\documentclass[11pt]{article}
\usepackage{mathrsfs}
\usepackage{amsmath}
\usepackage{amsfonts}
\usepackage{amssymb, amsmath, cite}
\usepackage{color}

\usepackage{mathrsfs}
\usepackage{dsfont}
\usepackage{amsthm}
\usepackage{mathrsfs}
\usepackage{amsmath}
\usepackage{amsfonts}
\usepackage[colorlinks,linkcolor=black,anchorcolor=blue,citecolor=blue]{hyperref}
\usepackage{amssymb, amsmath, cite}

\newcommand\be{\begin{equation}}
\newcommand\ee{\end{equation}}
\newcommand\ber{\begin{eqnarray}}
\newcommand\eer{\end{eqnarray}}
\newcommand\berr{\begin{eqnarray*}}
\newcommand\eerr{\end{eqnarray*}}
\newcommand\Om{\Omega}\newcommand\lm{\lambda}
\newcommand{\om}{{\omega}}

\newcommand{\dd}{\mathrm{d}}
\newcommand\e{\mathrm{e}}
{\setlength\arraycolsep{2pt}
\newcommand\nn{\nonumber}

\newcommand\bea{\begin{eqnarray}}
\newcommand\eea{\end{eqnarray}}
\newcommand\lb{\label}
\newcommand\eq{\eqref}
\newcommand{\rb}{r_{\mbox{\tiny B}}}
\newcommand{\xb}{x_{\mbox{\tiny B}}}
\newcommand{\ii}{\mathrm{i}}
\newcommand{\W}{{\cal W}}

\usepackage{graphicx}
\setlength{\baselineskip}{18pt}{\setlength\arraycolsep{2pt}

\begin{document}

\title{Yang--Mills Monopoles in \\Extremal Reissner--Nordstr\"{o}m Black Hole Metric}

\author{Xiangqin Zhang\\School of Mathematics and Statistics\\ Henan University\\
 Kaifeng, Henan 475004, PR China\\and\\
Henan University of Animal Husbandry and Economy\\ Zhengzhou, Henan 450044, PR China
\\\\
 Yisong Yang\footnote{Email address for correspondence: yisongyang@nyu.edu}\\Courant Institute of Mathematical Sciences\\New York University\\New York, NY 10012, USA}

\date{}

\maketitle

\begin{abstract}
We show that the closed-form monotone solution linking two different vacuum states, 
for the exterior Yang--Mills wave equation over the extremal Reissner--Nordstr\"{o}m spacetime found in a recent work of Bizo\'{n} and Kahl,
is the unique solution to an associated domain-wall type energy minimization problem. We also derive the accompanying interior Yang--Mills 
wave equation within the same formalism. We obtain a few more closed-form solutions, regular and singular, among other solutions
of oscillatory behavior, and discuss several interesting features of the solutions based on some energy consideration.

\medskip

\medskip

\medskip

\end{abstract}

\section{Introduction}
\setcounter{equation}{0}

The concept of magnetic monopoles was first conceived by P. Curie \cite{Curie} and later theoretically formulated by Dirac \cite{Dirac}
based on electromagnetic duality of the Maxwell theory. A remarkable feature of this concept is that its existence could lead to an
explanation why electric charges are multiples of a unit charge \cite{Dirac,GO,Pre,Ryder,Y}. Such a charge quantization phenomenon was
then extended by Schwinger \cite{Sch} to dyons, namely, particles of both electric and magnetic charges. In modern development,
thought experiments with monopoles immersed in a type-II superconductor led 
Mandelstam \cite{Man1,Man2}, Nambu \cite{Nambu}, and t' Hooft \cite{tH2,tH2}
to come up with a linear confinement mechanism for 
monopoles and their colored, non-Abelian or the Yang--Mills, field theory versions, which has propelled enormous activities and progress towards
an understanding \cite{SY1,SY2} of the unsettled puzzle of quark confinement \cite{Gre}. 
Like that in the classical Maxwell theory, a monopole in the pure Yang--Mills theory, for example, the celebrated Wu--Yang monopole
\cite{Actor,GO}, is of infinite energy \cite{JT}. There are several well-recognized paths which may be taken to overcome this infinity or divergence problem. These
include adopting the model of the Born--Infeld nonlinear electrodynamics \cite{BI1,BI2}, coupling the Yang--Mills fields with matter
fields \cite{Actor,GO,JZ,Pol,Pre,tH,Tyu,Wein,ZY}, namely, 
considering the Yang--Mills--Higgs theory, and hosting the Yang--Mills fields over curved spacetime backgrounds due to gravity. In the context of this last approach, if a gravitational black hole is present 
and fields are considered outside the event horizon so that possible singularities leading to energy divergence are concealed inside the horizon, as in the scenario
of the cosmic censorship hypothesis \cite{MTW,Wald}, the total energy of the fields outside the horizon could remain finite, as a result.
The present study belongs to this category of investigation. Specifically our work is motivated by the recent paper of Bizo\'{n} and Kahl \cite{BK}
who considered spherically symmetric $SU(2)$ Yang--Mills fields in the exterior of the event horizon of the extremal 
Reissner--Nordstr\"{o}m black hole and obtained a countable family of static regular solutions to the Yang--Mills equations,
given in terms of a scalar field sequence $\{W_n\}$, say, among which the first two, $W_1$ and $W_2$, are
in closed forms. These solutions resemble the Bartnick--McKinnon solutions \cite{BM,S} featured as soliton-like
configurations connecting vacuum states at infinity and 
oscillating about the unit-charge monopole state in local domains.
The main result of our present work is to show that, viewed as a connecting orbit between different vacuum states which passes
through the unit-charge monopole state midway, $W_1$ is the unique global energy minimizer of the problem. The solution $W_2$,
however, enjoys a different signature -- it exhibits itself as a local energy maximizer in a well-formulated sense. 
Besides, we shall also obtain a new closed-form solution which has two singular points (in fact, two singular shells), near which
energy blow-up takes place. Furthermore, we adapt the formalism of \cite{BK} to consider interior monopoles (inside the event 
horizon) and derive a governing wave equation which may be viewed as a companion of the wave equation of Bizo\'{n}--Kahl \cite{BK}
for exterior monopoles. For this interior monopole equation, we obtain three closed-form solutions, similar to $W_1$, $W_2$, and
the third singular solution to the exterior monopole equation described above, among which two are regular everywhere away from
the curvature singularity of the Reissner--Nordstr\"{o}m spacetime and the third one is singular at one point (in fact, one shell).
It is noted that this interior monopole equation possesses a family of oscillatory solutions as the exterior equation, as well.

The content of the rest of the paper is outlined as follows. In Section 2, we formulate the problems to be studied with
a review of the Yang--Mills fields over an extremal Reissner--Nordstr\"{o}m spacetime and the exterior monopole equation derived
in \cite{BK}, followed with an extension of such a formulation to get the  interior, `accompanying', monopole equation. Subsequently,
we focus on static equations.
In Section 3, we review the two closed-form regular solutions, $W_1$ and $W_2$, obtained in \cite{BK}, and present a third, singular, one, for
the exterior equation. We obtain the exact values of the energies of these two regular solutions to be further explored in later sections. We then present two closed-form regular solutions and a third, singular, one, for the interior equation, similar to those
for the exterior equation. In Section 4, we study several issues of the exterior solutions based on some energy consideration. 
In particular, we formulate a domain-wall type minimization problem that associates a distinguishing significance to $W_1$ as 
the unique global minimizer of the problem.  We also present a virial identity which displays an energy partition relation 
for finite-energy solutions. In Section 5, we first prove the existence of a solution to the domain-wall minimization problem 
formulated in Section 4. Note that, due to the behavior of the weight function in the energy functional, it is not a trivial matter to  
preserve the asymptotic limit of the weak limit of a minimizing sequence. Fortunately, such a difficulty may be overcome by a local
convergence method leading to realizing the weak limit as a monotone solution to the exterior equation and then by an energy 
comparison. We next prove the uniqueness of a monotone solution to the exterior equation. Since the energy functional is not
convex, such a uniqueness result is usually not ensured \cite{Evans} and seems surprising in the present context. Consequently, we recognize that $W_1$
is actually this energy minimizer whose energy calculated in Section 3 is thus the energy minimum of the formulated problem.
We also illustrate that $W_2$ is a local energy maximizer among a class of testing configurations, which seems to
suggest that, energetically, the closed-form solution $W_2$ of the exterior monopole equation has
a certain `significance' associated as well. In Section 6, we briefly comment on how to obtain oscillatory solutions for both exterior
and interior monopole equations using their polar-variable representations.

\section{Yang--Mills monopole equations in Reissner--Nordstr\"{o}m metric}

Consider the Reissner--Nordstr\"{o}m metric \cite{MTW,Wald}
\be\lb{2.1}
\dd s^2=-\left(1-\frac{2M}r+\frac{Q^2}{r^2}\right)\,\dd t^2+\left(1-\frac{2M}r+\frac{Q^2}{r^2}\right)^{-1}\,\dd r^2
+r^2(\dd\theta^2+\sin^2\theta\dd\phi^2),
\ee
describing a charged black hole of mass $M$ and charge $Q$ in the Schwarzschild spherical coordinates. At the critical situation
\be\lb{2.2}
M=Q,
\ee
the metric \eq{2.1} assumes the form 
\be\lb{2.3}
\dd s^2=-\frac{(r-M)^2}{r^2}\dd t^2+\frac{r^2}{(r-M)^2}\,\dd r^2+r^2 (\dd\theta^2+\sin^2\theta\dd\phi^2).
\ee
known as the extremal Reissner--Nordstr\"{o}m black hole metric with its event horizon at $r=M$. We will be interested in both
the exterior region, $r>M$, and interior one, $r<M$, possibly extended to its curvature singularity at $r=0$ as exhibited by the
associated Kretschmann invariant given by
\be\lb{xxK}
K=R_{\alpha\beta\mu\nu}R^{\alpha\beta\mu\nu}=\frac{8M^2}{r^8}(7M^2-12Mr+6r^2),
\ee
composed from the Riemann tensor induced by the gravitational metric form \eq{2.1}.

\subsection{Exterior region}

We first consider the space region  exterior to the event horizon, $r>M$. As in \cite{BK}, we use the variables
\be
\tau=\frac t{4M},\quad x=\ln\left(\frac rM-1\right),
\ee
to recast \eq{2.3} into
\bea
\dd s^2&=&\Om(x)\left(-\dd\tau^2+C^4(x)(\dd x^2+\dd\theta^2+\sin^2\theta\,\dd\phi^2)\right),\lb{xx2.5}\\
\Om(x)&=&\frac{16 M^2}{(1+\e^{-x})^2}\quad C(x)=\cosh\left(\frac x2\right).\lb{xx2.6}
\eea
Thus the exterior region in terms of the radial variable, $r>M$, is converted in terms of the variable $x$ to the full line $-\infty<x<\infty$, such that the `midway' spot, $x=0$, where the conformal factor $C(x)$ is minimized, corresponds to the classical Schwarzschild radius
\be
r_{\mbox{\small s}}=2M,
\ee
which will have a specific meaningfulness in our study.

We consider the $SU(2)$ Yang--Mills theory over such a spacetime.  Use $\sigma_a$ ($a=1,2,3$) to denote the Pauli matrices
and set
\be
t_a=\frac{\sigma_a}{2\ii},\quad a=1,2,3.
\ee
Then  $\{t_a\}$ generates $SU(2)$ and satisfies the commutator relation $[t_a,t_b]=\epsilon_{abc}t_c$. As in \cite{BK}, we choose the gauge field to be the 1-form
\be\lb{xx2.8}
A=W(\tau,x)\,\omega+t_3\cos\theta\,\dd\phi,\quad \omega=t_1\,\dd\theta+t_2\sin\theta\,\dd\phi,
\ee
where $W$ is a scalar field. Then we compute to get the Yang--Mills curvature field
\bea\label{x2.6}
F&=&\dd A+A\wedge A\nn\\
&=&t_1(W_\tau\,\dd\tau+W_x\,\dd x)\wedge\dd\theta+t_2\sin\theta (W_\tau\,\dd\tau
+W_x\,\dd x)\wedge\dd\phi
+t_3\sin\theta(W^2-1)\,\dd\theta\wedge\dd\phi.\quad
\eea
Therefore, with $x^0=\tau,x^1=x,x^2=\theta,x^3=\phi$ and $F=F^a_{\mu\nu}t_a\,\dd x^\mu\wedge\dd x^\nu$, we can read 
\eqref{x2.6} to obtain the nonzero and independent components of the 2-form $F$ to be
\be\label{x2.7}
F^1_{02}=\frac12W_\tau,\quad F^1_{12}=\frac12W_x,\quad F^2_{03}=\frac12\sin\theta W_\tau,\quad F^2_{13}=\frac12\sin\theta W_x,\quad F^3_{23}=\frac12
\sin\theta(W^2-1).
\ee
On the other hand, if we rewrite line element \eqref{xx2.5}--\eq{xx2.6} as $\dd s^2=g_{\mu\nu}\,\dd x^\mu\dd x^\nu$, then
the Yang--Mills action density reads
\bea\lb{xx2.11}
{\cal L}&=&\frac14 \,F^a_{\mu\nu}g^{\mu\alpha}g^{\nu\beta}F^a_{\alpha\beta}\sqrt{|\det(g_{\alpha\beta})|}\nn\\
&=&\left(-\frac{C^2}4 W_\tau^2+\frac1{4C^{2}} W_x^2+\frac1{8 C^{2}}(1-W^2)^2\right)\sin\theta,
\eea
in view of \eq{x2.7}. Hence we can rewrite the Yang--Mills action as
\be
L=\int{\cal L}\,\dd^4 x=2\pi \int{\cal L}_W\,\dd x\dd \tau,
\ee
where
\be\lb{xx2.13}
{\cal L}_W=-\frac{C^2}2 W_\tau^2+\frac1{2C^{2}} W_x^2+\frac1{4 C^{2}}(1-W^2)^2,
\ee
as obtained in \cite{BK}, so that the Yang--Mills equation is the Euler--Lagrange equation of \eq{xx2.13}:
\be\lb{xx2.14}
W_{\tau\tau}=C^{-2}(C^{-2} W_x)_x+C^{-4} W(1-W^2),
\ee
whose static limit reads
\be\lb{xx2.15}
W''-\tanh\left(\frac x2\right) W'+W(1-W^2)=0,
\ee
where and in the sequel, we use the notation $W'=W_x$, etc., interchangeably.

\subsection{Interior region}
We next consider the region that is interior to the event horizon, $0<r<M$. Thus, using the variables
\be\lb{xx2.17}
\tau=\frac t{4M},\quad x=\ln\left(1-\frac rM\right),\quad 0<r<M,
\ee
we have $-\infty<x<0$ and
we see that the line element \eq{2.3} becomes
\bea
\dd s^2&=&\Om(x)\left(-\dd\tau^2 +S^4(x)(\dd x^2+\dd\theta^2+\sin^2\theta\dd\phi^2)\right),\label{x2.16}\\
\Om(x)&=&\frac{16 M^2}{(1-\e^{-x})^2},\quad S(x)=\sinh\left(\frac x2\right).\lb{x2.17}
\eea
Therefore, from \eq{xx2.8} and \eq{x2.6},  we analogously obtain the effective Yang--Mills action density
\be
{\cal L}_W=-\frac{S^2}2 W_\tau^2+\frac1{2 S^2} W_x^2+\frac1{4 S^2}(1-W^2)^2,
\ee
over the spatial interval $-\infty<x<0$, with the associated Euler--Lagrange equation
\be\label{x2.12}
W_{\tau\tau}={S^{-2}}(S^{-2}W_x)_x+{S^{-4}}(1-W^2)W,
\ee
whose static limit reads
\be
W''-\coth\left(\frac x2\right) W'+W(1-W^2)=0.\label{x2.21}
\ee

The equations \eq{x2.12} and \eq{x2.21} are a new pair of equations, accompanying \eq{xx2.14} and \eq{xx2.15}.

\section{Exact solutions}
\setcounter{equation}{0}

In \cite{BK}, the following two nontrivial exact solutions to the exterior Yang--Mills monopole equation \eq{xx2.15} are obtained:
\be\lb{3.1}
W_1(x)=\tanh\left(\frac x2\right),\quad W_2(x)=\frac{2\cosh x-2-\sqrt{6}}{2\cosh x+4+3\sqrt{6}}.
\ee
Here we find that there is a third exact solution,
\be\lb{3.2}
W_3(x)=\frac{2\cosh x-2+\sqrt{6}}{2\cosh x+4-3\sqrt{6}},
\ee
which is apparently singular at
\be
x_\pm=\pm\ln\left(\frac{3\sqrt{6}}2+2\sqrt{3}-2-\frac{3\sqrt{2}}2\right)\approx \pm 1.104268225,
\ee
giving rise to the corresponding singular radii
\be
r_{\pm}=\left(1+\e^{x_\pm}\right)M,
\ee
in the original radial variable, which are distributed about the Schwarzschild radius $r_{\mbox{\small s}}$ following
\be
M<r_-<r_{\mbox{\small s}}=2M<r_+,\quad \quad \frac{r_-+r_+}2=\left(1+\cosh(x_+)\right)M>r_{\mbox{\small s}}.
\ee

For the interior Yang--Mills monopole equation \eq{x2.21}, we likewise find the following three nontrivial exact solutions:
\bea
w_1(x)&=&\coth\left(\frac x2\right),\lb{3.4}\\
w_2(x)&=&\frac{2\cosh x+2-\sqrt{6}}{2\cosh x-4+3\sqrt{6}},\lb{3.5}\\
w_3(x)&=&\frac{2\cosh x+2+\sqrt{6}}{2\cosh x-4-3\sqrt{6}},\lb{3.6}
\eea
for $-\infty<x<0$. The solutions $w_1$ and $w_2$ given in \eq{3.4} and \eq{3.5}, respectively, are regular but $w_3$ in \eq{3.4} is
 singular at
\be\lb{3.7}
x_0=-\ln\left(2[1+\sqrt{3}]+\frac3{\sqrt{2}}[1+\sqrt{3}]\right),
\ee
whereby the irrelevant positive root is discarded.

We now compute the energies of these solutions.

Recall that the energy-momentum tensor induced from \eq{xx2.11} is
\be\lb{3.8}
T_{\mu\nu}=-F^a_{\mu\alpha}g^{\alpha\beta}F^a_{\nu\beta}+\frac14 g_{\mu\nu}(F^a_{\gamma\delta}g^{\gamma\alpha}g^{\delta\beta}F^a_{\alpha\beta}),
\ee
which via ${\cal H}=T^0_0=g^{00}T_{00}$ gives rise to the  Hamiltonian energy density associated with \eq{xx2.5}--\eq{xx2.6} and 
\eq{x2.7} to be
\be
{\cal H}_W=\frac{C^2}2 W_\tau^2+\frac1{2C^{2}} W_x^2+\frac1{4 C^{2}}(1-W^2)^2.
\ee
Thus, evaluating the static total energy
\be\lb{3.10}
E(W)=2\pi\int_{-\infty}^\infty\left(\frac1{2C^{2}} W_x^2+\frac1{4 C^{2}}(1-W^2)^2\right)\,\dd x,
\ee
we obtain the exact value
\be\lb{3.11}
E(W_1)=\frac{8\pi}5,
\ee
for the solution $W_1$ given in \eq{3.1}. For the solution $W_2$ given in \eq{3.1}, utilizing MAPLE, we obtain the exact value
\be\lb{3.12}
E(W_2)=2\pi \frac{(12198704\sqrt{2}\sqrt{3}+30388736)\tanh^{-1}\left(\frac5{3\sqrt{2}+\sqrt{3}}\right)+15707196\sqrt{2}+13222288\sqrt{3}}{35368572\sqrt{3}+43079049\sqrt{2}},
\ee
which is approximately $2\pi (0.9664005272)$. These two exact results are consistent with those found in \cite{BK} based on numerical evaluation. Not surprisingly, the singular solution $W_3(x)$ is of infinite energy. In other words, this solution
behaves singularly in both point-wise and energy-wise manners at $x_\pm$.

We now turn our attention to the interior solutions listed in \eq{3.4}--\eq{3.6}. In view of  \eq{3.8}, \eq{x2.7}, and
\eq{x2.16}--\eq{x2.17}, we get the Hamiltonian energy density
\be\lb{3.13}
{\cal H}_W=\frac{S^2}2 W_\tau^2+\frac1{2S^{2}} W_x^2+\frac1{4 S^{2}}(1-W^2)^2,
\ee
which renders the total energy for a static interior solution in the full region $0<r<M$ to be
\be\lb{3.14}
E=2\pi\int_{-\infty}^0\left(\frac1{2S^{2}} W_x^2+\frac1{4 S^{2}}(1-W^2)^2\right)\,\dd x.
\ee
All the solutions given in \eq{3.4}--\eq{3.6} are of infinity energy due to the divergence of the energy density \eq{3.13} as a consequence of the curvature singularity of the Reissner--Nordstr\"{o}m black hole spacetime at $r=0$
displayed by \eq{xxK}. Nevertheless, we may assume
that we are in a situation that gravity is generated from a bulk of matter which occupies a finite region within the event horizon,
\be
r\leq \rb,\quad 0<\rb<M.
\ee
Then we can focus on the `regularized' region
\be\lb{3.16}
-\infty<x<\xb<0,\quad \xb=\ln\left(1-\frac \rb{M}\right),\quad \rb\leq r<M,
\ee
instead. It is seen that the solutions \eq{3.4} and \eq{3.5} are of finite energies over \eq{3.16} for any $0<\rb<M$ and the
solution \eq{3.6} is of finite energy only when $\xb$ is below its singular point $x_0$ spelled out in \eq{3.7}. Thus, we are led
to the finite-energy condition
\be
\xb<-\ln\left(2[1+\sqrt{3}]+\frac3{\sqrt{2}}[1+\sqrt{3}]\right)\approx - 2.421226122,
\ee
or equivalently, 
\be
\rb>\left(\frac{(3+3\sqrt{3})\sqrt{2}+4\sqrt{3}+2}{(3+3\sqrt{3})\sqrt{2}+4\sqrt{3}+4}\right)M\approx (0.9111873445)M.
\ee
Over a given interval \eq{3.16}, the energy of the solution \eq{3.4} is generically much greater than that of the solution \eq{3.5},
since the former is singular at its right boundary point, $x=0$, and the latter is everywhere regular. As a comparison, we list
our calculation of the energy
\be\lb{3.14}
E(W)=2\pi\int_{-\infty}^{\xb}\left(\frac1{2S^{2}} W_x^2+\frac1{4 S^{2}}(1-W^2)^2\right)\,\dd x,
\ee
for $\xb=-1$ for the solutions $w_1$ and $w_2$, given in \eq{3.4} and \eq{3.5}, respectively, as follows:
\be
E(w_1)=2\pi(3.273953196),\quad E(w_2)=2\pi(0.3220989496).
\ee

\section{Energy consideration of exterior monopoles}
\setcounter{equation}{0}

We now investigate the problem whether there is a nontrivial energetically stable exterior monopole solution linking two vacuum states,
$W=\pm1$, characterized by
the boundary condition $W(-\infty)=\pm1$ and $W(\infty)=\pm1$. It is clear that
the problem of linking the same vacuum state with $W(-\infty)=W(\infty)=\pm1$ is not well posed without some further elaboration 
since the vacuum states themselves minimize the energy and may be approached arbitrarily by nontrivial field configurations.
Thus it remains to consider the problem of linking two different vacuum states, say,
\be\lb{4.1}
\eta=\inf\{ E(W)\,|\, W(-\infty)=-1,W(\infty)=1\},
\ee
and ask whether this minimization problem has a solution, where $E(W)$ is as given in \eq{3.10}. This is typically a domain wall
problem \cite{CCY,CY,CY2,R,V}, interpolating two 
ground state domains, $W=\pm1$. However, it is not hard to see that this problem is not well posed either.

In fact, let $u(x)$ be a smooth function satisfying 
\be
u(x)=-1,\quad x\leq -1;\quad u(x)=1,\quad x\geq1.
\ee
Then, using $u_{(a)}=u(x-a)$ as a testing function, we have
\bea
E(u_{(a)})&=&\pi\int_{a-1}^{a+1}\frac1{C^2(x)}\left((u'_{(a)})^2+\frac12(u^2_{(a)}-1)^2\right)\,\dd x\nn\\
&\leq&K_0\int_{a-1}^{a+1}\frac{\dd x}{C^2(x)}\to0\quad \mbox{as }|a|\to\infty.
\eea
Here $K_0>0$ is a constant depending on the properties of $u$ over $[-1,1]$. As a consequence, this shows
the quantity $\eta$ in \eq{4.1} is zero which
is therefore not attainable.

\begin{figure}[h]
\begin{center}
\includegraphics[height=7cm,width=9cm]{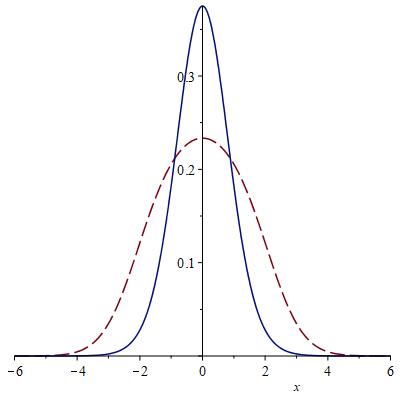}
\caption{Plots of the energy distributions of the solutions linking two different vacuum states and the same vacuum state
and the unit-charge monopole state viewed as an intermediate state. The former is depicted by a solid-line curve showing the concentration of energy at the spot where the solution passes through the
intermediate state. The latter is drawn by a dash-line curve exhibiting the concentration of energy in the middle of the two spots
where the solution passes through the intermediate state. Energetically both solutions demonstrate themselves as domain-wall solitons.}
\label{F1}
\end{center}
\end{figure}

On the other hand, since the boundary condition in \eq{4.1} dictates that
the unit magnetic charge monopole configuration, $W=0$, is to occur somewhere as an intermediate state, we may conveniently impose the following additional `intermediate state condition'
\be
W(0)=0.
\ee
It is interesting that this condition implies that the potential part of the energy density
\be\lb{4.5}
{\cal H}_W=\frac1{2C^{2}(x)} W_x^2+\frac1{4 C^{2}(x)}(1-W^2)^2
\ee
is maximized at the spot where the intermediate state occurs, $x=0$. In Figure \ref{F1}, we plot the energy density \eq{4.5} for
the solutions $W_1$ and $W_2$ given in \eq{3.1} as an illustration. It is interesting to notice that the former peaks sharply at $x=0$
where the solution passes the intermediate state and the latter also peaks at the same spot which is not where the solution passes
the intermediate state but is the midpoint between the two spots,
\be
x^{\pm}_0=\ln\left(1+\frac{\sqrt{6}}2\pm\frac{\sqrt{6+4\sqrt{6}}}2\right),
\ee
 where the solution passes the intermediate state. Such
a property of energy concentration is typical for solitons in field theories.

These pictures coincide with what we know about a domain wall soliton configuration in general. 

Thus, we are led to modifying \eq{4.1} into the minimization problem 
\be\lb{4.6}
\eta_0\equiv \inf\{ E(W)\,|\, W(-\infty)=-1,W(\infty)=1, W(0)=0\}.
\ee

We shall show that $W_1$ given in \eq{3.1} is the unique solution to this problem. Consequently, as a by-product, we have
\be
\eta_0=\frac{8\pi}5,
\ee
in view of \eq{3.11}.

Let $W$ be a finite-energy critical point of the energy functional \eq{3.10}. Set $W_\lm(x)=W(\lm x)$. Then,
from $\frac{\dd}{\dd\lm}E(W_\lm)|_{\lm=1}=0$, we get  the virial identity
\be
\int_{-\infty}^{\infty}\frac1{C^2(x)}\left(\left[1+x\tanh\left(\frac x2\right)\right] W_x^2+\frac x2\, \tanh\left(\frac x2\right)(W^2-1)^2\right)\,\dd x=\int_{-\infty}^\infty\frac{(W^2-1)^2}{2C^2(x)}\,\dd x,
\ee
which is an energy partition relation. This identity demonstrates that the `potential energy' is much greater than the `elastic energy'
for an exterior monopole. For example, for the exact solution $W=W_1$ in \eq{3.1}, we have
\be
\int_{-\infty}^\infty\frac{(W^2-1)^2}{4C^2(x)}\,\dd x=\frac8{15}, \quad \int_{-\infty}^{\infty}\frac{W_x^2}{2C^2(x)}\,\dd x=\frac4{15},
\ee
and, for $W=W_2$ given in \eq{3.1}, these quantities are approximately $0.8765154575$ and $0.08988507022$, respectively.

\section{Existence and uniqueness of exterior energy minimizer}
\setcounter{equation}{0}

The symmetry of the functional \eq{3.10} indicates that the problem \eq{4.6}  amounts to considering the minimization problem
\be\lb{5.1}
I_0\equiv \inf\left\{I(W)\equiv \int_0^\infty \frac1{C^2(x)}\left((W')^2+\frac12(W^2-1)^2\right)\,\dd x\,\bigg|\,W(0)=0,W(\infty)=1\right\},
\ee
over the set of admissible functions which are absolutely continuous on all compact subintervals of the half line $[0,\infty)$.

Let $\{\W_n\}$ be a minimizing sequence of \eq{5.1} satisfying
\be\lb{5.2}
\lim_{n\to\infty}I(\W_n)=I_0;\quad I(\W_n)\leq I_0+\frac1n,\quad n=1,2,\dots.
\ee
By modifying the sequence $\{\W_n\}$ if necessary, we may assume that $\{\W_n\}$ enjoys the property that each function $\W_n$ satisfies $0\leq \W_n(x)\leq 1$. 
Furthermore, for each $\W_n$, we may modify it such that it minimizes the
partial functional
\be
I_n(W)\equiv \int_0^n\frac1{C^2(x)}\left((W')^2+\frac12(W^2-1)^2\right)\,\dd x,
\ee
among functions fulfilling the boundary condition $W(0)=0, W(n)=\W_n(n)$ since the functional is differentiable and
weakly lower semicontinuous over the Sobolev space $W^{1,2}(0,n)$, say (cf. Theorem 2 on page 448 in Evans\cite{Evans} 
specifically as well as the studies in \cite{GT,Tyu,ZY} in other contexts).
Therefore $\W_n$ when restricted to the interval $[0,n]$
is a critical point of $I_n$. As a consequence, we have
\be\lb{5.4}
\int_0^n\frac1{C^2(x)}\left( \W'_n w'+(\W_n^2-1)\W_n w\right)\,\dd x=0,\quad \forall w\in C_0^1(0,n).
\ee

Extracting a suitable subsequence if necessary, for example, using a diagonal subsequence argument, we may assume that $\{\W_n\}$ is weakly convergent 
in the Sobolev space $W^{1,2}(0,a)$, with the weighted measure $ C^{-2}(x)\,\dd x$, for {\em any} $a>0$. Let ${\W}$ denote
such a weak limit which is well defined over $[0,\infty)$.
Fix $a>0$.
Neglecting the first few terms of the sequence if necessary, we may also assume $n>a$. Thus \eq{5.4} gives us
\be\lb{5.5}
\int_0^a\frac1{C^2(x)}\left( \W'_n w'+(\W_n^2-1)\W_n w\right)\,\dd x=0,\quad \forall w\in C_0^1(0,a).
\ee
Using the compact embedding from the Sobolev space into the space $C[0,a]$ and letting $n\to\infty$ in \eq{5.5}, we arrive at
\be\lb{5.6}
\int_0^a\frac1{C^2(x)}\left({\W}' w'+({\W}^2-1){\W} w\right)\,\dd x=0,\quad \forall w\in C_0^1(0,a),
\ee
which leads us to conclude by the standard elliptic regularity theory \cite{GT,LU} that ${\W}$ is a classical solution to the Euler--Lagrange equation of the functional $I$. Hence
$\W$ satisfies
\be\lb{5.7}
\left(\frac{\W'}{C^2(x)}\right)'+\frac1{C^2(x)}(1-\W^2)\W=0,\quad \W(0)=0.
\ee
Since $\{\W_n\}$ is a sequence with values confined in $[0,1]$, so is its limit $\W$. However, since $\W$ satisfies 
\eq{5.7}, we see that $\W\equiv0$ or $0<\W(x)<1$ for $x>0$ because $\W\equiv0$ and $\W\equiv1$ are two equilibria 
of the differential equation in \eq{5.7}.
We now show that the former does not occur.

In fact, by using weak convergence, we have
\be\lb{5.8}
I_a(\W)\leq \liminf_{n\to\infty} I_a(\W_n)\leq \lim_{n\to\infty} I(\W_n)=I_0,
\ee
for any $a>0$. Thus, letting $a\to\infty$ in \eq{5.8}, we get
\be\lb{5.9}
I(\W)\leq I_0.
\ee
Recall the result
\be\lb{5.10}
I_0\leq I\left(\tanh\,\frac x2\right)=\frac45.
\ee
Besides, we also have $I(0)=1$. Hence $\W\not\equiv0$. Consequently, $\W$ satisfies $0<\W(x)<1$ for $x>0$. 

Since $\W$ is bounded, we deduce that there is a sequence $\{x_n\}$, $x_n\to\infty$ as $n\to\infty$, such that
$\W'(x_n)\to0$ as $n\to\infty$. Using this result as the boundary condition and integrating the differential equation in \eq{5.7}, we have
\be\lb{5.12}
\frac {\W'(x)}{C^2(x)}=\int^\infty_x \frac1{C^2(y)}(1-\W^2(y))\W(y)\,\dd y.
\ee
Hence $\W'(x)>0$ for $x>0$.
In particular,
\be\lb{5.11}
\lim_{x\to\infty}\W(x)=L,
\ee
for some $L\in (0,1]$.
Thus, if $L<1$ in \eq{5.11}, we infer from \eq{5.12} that
\be
\lim_{x\to\infty}\W'(x)=(1-L^2)L>0,
\ee
 which is false. Therefore $L=1$
in \eq{5.11}. This establishes that $\W$ lies in the admissible space of the problem \eq{5.1}. Thus $I(\W)\geq I_0$.
Combining this result with \eq{5.9}, we arrive at
\be
I(\W)=I_0.
\ee
In other words, the existence of a solution to the minimization problem \eq{5.1} follows.

We next prove that, actually,
\be\lb{5.14}
\W(x)=\tanh\left(\frac x2\right).
\ee
Thus, in particular, equality in \eq{5.10} holds. We achieve this goal by showing that a positive solution to the initial value
problem \eq{5.7}, with a limiting value at infinity, is unique.

In fact, let $\W_1$ and $\W_2$ be two such solutions. Then $w=\W_1-\W_2$ satisfies
\be\lb{5.15}
\left(\frac {w'}{C^2(x)}\right)'+\frac w{C^2(x)}(1-\W_1^2-\W_2^2-\W_1 \W_2)=0.
\ee
On the other hand, for any differentiable function $h$, we have
\bea\lb{5.16}
&&C^2(x)\left(\frac{\W_1 \W'_1 h^2}{C^2(x)}\right)'+{\W_1^2 (h')^2}+{(\W_1\W_2+\W_2^2)\W_1^2 h^2}\nn\\
&&=([\W_1 h]')^2-(1-\W_1^2-\W_2^2-\W_1\W_2)(\W_1 h)^2.
\eea
Now substituting $h=\frac w{\W_1}$ into \eq{5.16}, we have
\bea\lb{5.17}
&&\frac{(w')^2}{C^2}-\frac1{C^2}(1-\W_1^2-\W_2^2-\W_1 \W_2)w^2\nn\\
&&=\left(\frac{\W_1' w^2}{C^2 \W_1}\right)'+
\frac{(w' \W_1-w\W_1')^2}{C^2}+\frac{(\W_1\W_2+\W_2^2)w^2}{C^2}.
\eea
Moreover, using the Cauchy--Kovalevskaya theorem, we know that the solution of \eq{5.7} has the asymptotic form
\be\lb{5.18}
\W(x)=bx+ cx^3+\mbox{O}(x^5),\quad x\sim 0.
\ee
As a consequence of \eq{5.18}, we may integrate \eq{5.17} and drop the resulting boundary terms to arrive at
\bea\lb{5.19}
&&\int_0^\infty\left(\frac{(w')^2}{C^2}-\frac1{C^2}(1-\W_1^2-\W_2^2-\W_1 \W_2)w^2\right)\,\dd x\nn\\
&&=\int_0^\infty \frac1{C^2(x)}\left( {(w' \W_1-w\W_1')^2}+{(\W_1\W_2+\W_2^2)w^2}\right)\,\dd x.
\eea
However, by virtue of \eq{5.15}, we may rewrite the left-hand side of \eq{5.19} as
\bea\lb{5.20}
&&\int_0^\infty\left(\frac{(w')^2}{C^2}-\frac1{C^2}(1-\W_1^2-\W_2^2-\W_1 \W_2)w^2\right)\,\dd x\nn\\
&&=\left(\frac{ww'}{C^2}\right)_{0}^{\infty}-\int_0^\infty w\left(\left[\frac {w'}{C^2}\right]'+\frac w{C^2}(1-\W_1^2-\W_2^2-\W_1 \W_2)\right)\,\dd x=0.
\eea
Combining \eq{5.19} and \eq{5.20}, we conclude with $w\equiv 0$.

Hence the existence and uniqueness of an exterior energy-minimizing monopole solution linking different vacuum states at
$x\pm\infty$ and
passes through the unit-charge monopole state at $x=0$ is established which is the exact solution $W_1$ stated in \eq{3.1}
and found in \cite{BK} which determines the minimum energy, $E_{\min}=\frac{8\pi}5$,  as stated in \eq{3.11}.

It is interesting to note that a rich family of   even and odd solutions of \eq{xx2.15} linking the same and different vacuum
states is obtained in \cite{BK} which are characterized
by the number $n$ of zeros they possess. When $n=1$, their calculated energy coincides with our exact value; when $n\geq 2$, their calculated
energies are all significantly greater than that with $n=1$.

Motivated by the construction in this section and the even solutions with multiple zeros linking to the same vacuum state, $W(-\infty)
=W(\infty)=\pm1$, obtained in \cite{BK}, we may impose the following minimization problem
\be\lb{5.21}
\sigma_0\equiv\inf \{E(W)\,|\, W(-\infty)=W(\infty)=1, W(0)\leq0\},
\ee
over the admissible set of functions which are absolutely continuous on all compact subintervals of the real line. Thus, with
the function \eq{5.14} as a testing function, i.e., choosing $W(x)=|\tanh(\frac x2)|$ for all $x$, we have the estimate 
\be
\sigma_0\leq \frac{8\pi}5.
\ee

\begin{figure}[h]
\begin{center}
\includegraphics[height=7cm,width=9cm]{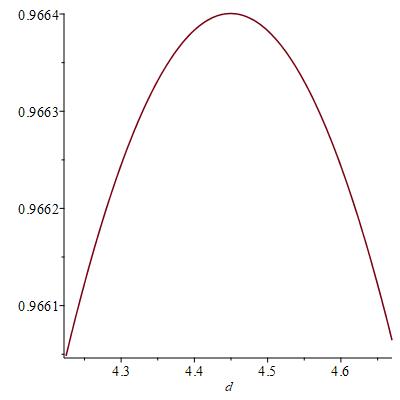}
\caption{A plot of the normalized energy $\frac{E(W_h)}{2\pi}$ against $d=2+3(2+\sqrt{6})h$ for $\frac16<h<\frac15$ such
that the critical height $h_0=\frac1{3+\sqrt{6}}$ corresponds to $d_0=\frac{12+5\sqrt{6}}{3+\sqrt{6}}\approx 4.45$
where the energy peaks with the signature value, $0.9664$, first obtained in \cite{BK}.}
\label{F2}
\end{center}
\end{figure}

With the problem \eq{5.21} in mind, we consider the exact solution $W_2$ given in \eq{3.1} whose value at $x=0$ is
\be
W_2(0)=-h_0,\quad h_0=\frac1{3+\sqrt{6}},
\ee
and we ask what this `height' $h_0$ means with regard to the problem \eq{5.21}. So we take the trial function
\be\lb{5.24}
W_h(x)=\frac{2\cosh x-d}{2\cosh x+4+3\sqrt{6}},\quad d\equiv  2+3(2+\sqrt{6})h,\quad 0<h<1,
\ee
which satisfies $W_h(0)=-h$ and is suggested by the form of $W_2(x)$ in \eq{3.1}. Hence we can compute the energy
$E(W_h)$. Surprisingly, we find that $E(W_h)$ is {\em maximized} at $h=h_0$ as illustrated in Figure \ref{F2}.

Therefore, the quantity $h_0$ may be regarded as a `canonical height' at which the energy confined over the profile family of functions of the
form \eq{5.24} is maximized and brings forth a critical point of the energy functional.

\section{Remarks on oscillatory solutions}
\setcounter{equation}{0}

In \cite{BK}, a family of solutions of \eq{xx2.15} with prescribed numbers of zeros are obtained. In this section, we revisit such
a construction of oscillatory solutions from the viewpoint of recasting the equation into a form of its `polar variable' representation
such that the appearance of multiple zeros of the solutions becomes somewhat more transparent as we now show. 

For our purpose, we use the following polar variable anzatz 
\be\lb{6.1}
W=\rho\cos\om,\quad W'=\rho \sin\om,
\ee
to represent a solution of \eq{xx2.15}. Thus, we see that the pair $(\rho,\om)$ satisfies the coupled system of equations
\bea
\rho'&=&\tanh\left(\frac x2\right)\,\rho\sin^2\om+\rho^3\cos^3\om\sin\om,\lb{6.2}\\
\om'&=&-1+\tanh\left(\frac x2\right)\,\sin\om\cos\om+\rho^2\cos^4\om.\lb{6.3}
\eea

To justify this representation, we observe in view of the uniqueness theorem for solutions of the initial value problems of ordinary
differential equations that if $\rho$ satisfying \eq{6.2} is positive at a point then it will stay positive where it exists. So, we may
impose the initial condition
\be\lb{6.4}
\rho(0)=a>0,\quad \om(0)=b.
\ee
Thus, in particular, to produce an even solution of \eq{xx2.15}, we  take  $b=0$, and an odd solution, we impose $b=\frac\pi2$. 
In the subsequent discussion, we fix $b$ and adjust $a$ to achieve multiple zeros for the field $W$. Accordingly, in view of
\eq{6.1}, such a goal may be achieved by making the angular variable $\om$ circulate through the full circle multiple times, 
in a sense to be made precise below.  

Indeed, we fix $b,K_0>0$. Then the theorem of continuous dependence of solutions on the initial data allows us to get
some number $a_0>0$ such that the solution to \eq{6.2}--\eq{6.4} satisfies
\be\lb{6.5}
0<\rho(x)\leq \frac12,\quad 0\leq x\leq K_0,
\ee
when $a\in(0,a_0]$.
Applying \eq{6.5} to \eq{6.3} and using $\sin\om\cos\om\leq\frac12$, we get
\be\lb{6.6}
\om'(x)\leq-\frac14,\quad 0\leq x\leq K_0.
\ee
Therefore, for any integer $n\geq1$, we may set $K_0=4n\pi$ (say). Then \eq{6.6} indicates that there are $n$ points
$x_1,\dots,x_n$ in $(0,K_0]$ such that
\be
\om(x_i)=\mbox{an odd multiple of }\, \frac\pi2,\quad i=1,\dots,n.
\ee
As a consequence of this observation and \eq{6.1}, we see that if $b$ is not an odd multiple of $\frac\pi2$, then $x_1,\dots,x_n$
are the zeros of $W$, and if $b$ is an odd multiple of $\frac\pi2$, then $b,x_1,\dots,x_n$ are the zeros of $W$.
Hence a construction of a local solution with a prescribed number of zeros is obtained. Note that such a solution may be even, odd, or
neither.

For the interior monopole equation \eq{x2.21}, if we use the polar representatiion \eq{6.1} with the flippled variable $x=-y$,
then we get the transformed governing system
\bea
\rho'&=&\coth\left(\frac y2\right)\,\rho\sin^2\om-\rho^3\cos^3\om\sin\om,\lb{6.8}\\
\om'&=&1+\coth\left(\frac y2\right)\,\sin\om\cos\om-\rho^2\cos^4\om,\lb{6.9}
\eea
where $0<y<\infty$ and $\rho'=\frac{\dd\rho}{\dd y}$, etc. Although $\coth(\frac y2)$ is unbounded as $y\to0$, 
corresponding to the curvature singularity, we may again restrict our attention to a region away from the singularity in order to
get solutions with multiple zeros in $W$. For example, it suffices to request $\coth(\frac y2)<2$ for our method here to work. Such
a condition translates into
\be
y>\ln 3.
\ee
In view of \eq{xx2.17}, this restriction gives us the interior region
\be
\frac{2M}3<r<M,
\ee
to accommodate oscillatory monopole solutions similar to those  in the exterior region.

In order to obtain a global solution with a prescribed number of zeros with desired asymptotic limits at infinity, $W(\pm\infty)=\pm1$,
additional work is needed for the choice of the initial data. Here we omit the details since much of the idea
along the line is presented in \cite{BK} mathematically and numerically. See also \cite{H,Mc,S}.

\medskip

The authors would like to thank an anonymous referee whose comments and constructive suggestions helped improve the presentation of
this paper.

\medskip

{\bf Data availability statement:} The data that supports the findings of this study are available within the article.

\end{document}